\documentclass[showpacs,aps,twocolumn]{revtex4}
\usepackage{epsfig}
\usepackage{graphicx}
\usepackage{amsmath,amssymb,amsfonts}
\usepackage{array}
\usepackage{url}
\usepackage{hyperref}
\usepackage{multirow}
\usepackage{float}
\usepackage{lineno}
\usepackage{xspace}
\usepackage[usenames,dvipsnames]{color}

\newcommand{\bef}{\begin{figure}}
\newcommand{\eef}{\end{figure}}
\newcommand{\bc}{\begin{center}}
\newcommand{\ec}{\end{center}}

\newcommand{\be}{\begin{equation}}
\newcommand{\ee}{\end{equation}}
\newcommand{\bea}{\begin{eqnarray}}
\newcommand{\eea}{\end{eqnarray}}

\def\ba{\begin{eqnarray}}
\def\ea{\end{eqnarray}}
\begin{document}
\title{Role of event multiplicity on hadronic phase lifetime and QCD phase boundary in
ultrarelativistic collisions at energies available at the BNL Relativistic Heavy Ion
Collider and CERN Large Hadron Collider}
\author{Dushmanta Sahu}
\author{Sushanta Tripathy}
\author{Girija Sankar Pradhan}
\author{Raghunath Sahoo\footnote{Corresponding Author Email: Raghunath.Sahoo@cern.ch}}
\affiliation{Discipline of Physics, School of Basic Sciences, Indian Institute of Technology Indore, Simrol, Indore 453552, INDIA}

\begin{abstract}
Hadronic resonances, having very short lifetime, like $\rm{K}^{*0}$, can act as useful probes to understand and estimate lifetime of hadronic phase in ultra-relativistic proton-proton, p--Pb and heavy-ion collisions. Resonances with relatively longer lifetime, like $\phi$ meson, can serve as a tool to locate the QGP phase boundary. We estimate a lower limit of hadronic phase lifetime in Cu--Cu and Au--Au collisions at RHIC, and in pp, p--Pb and Pb--Pb collisions at different LHC collision energies. Also, we obtain the effective temperature of $\phi$ meson using Boltzmann-Gibbs Blast-Wave function, which gives an insight to locate the QGP phase boundary. We observe that the hadronic phase lifetime strongly depends on final state charged-particle multiplicity, whereas the QGP phase and hence the QCD phase boundary shows a very weak multiplicity dependence. This suggests that the hadronisation from a QGP state starts at a similar temperature irrespective of charged-particle multiplicity, collision system and collision energy, while the endurance of hadronic phase is strongly dependent on final state charge-particle multiplicity, system size and collision energy.

\end{abstract}
\date{\today}
\maketitle

\section{Introduction}
\label{intro}
To reveal the nature of the QCD phase transition and to get a glimpse of how matter behaves at such extreme conditions of temperature and energy density, experiments like RHIC at BNL, USA and the LHC at CERN, Geneva, Switzerland have performed hadronic and heavy-ion collisions at ultra-relativistic energies. A deconfined state of quarks and gluons, also known as Quark Gluon Plasma (QGP), is believed to be produced for a very short lifetime in heavy-ion collisions in these experiments. In the QGP phase, the relevant degrees of freedom are partons: quarks and gluons. In a hadronic phase, the composite objects like mesons and baryons are the degrees of freedom \cite{Aoki:2006we}. QGP is governed by Quantum Chromodynamics (QCD) and it is the result of a first order/cross-over phase transition from normal nuclear matter consisting of mesons and baryons~\cite{Harris:1996zx,BraunMunzinger:2007zz}. Traditionally, it was believed that for small collision systems, the spacetime evolution 
could be different than heavy-ion collisions. This means, in heavy-ion collisions, where the formation of a QGP phase is expected, may undergo various processes like- pre-equilibrium of partons, a QGP phase, a possible mixed phase of partons and hadrons during hadronisation, hadronic phase and finally freeze-out. Contrary to this, in hadronic collisions, where usually one doesn't expect the formation of a partonic medium, the system may undergo multiparticle production in the final state, without having a QCD phase transition. The Large Hadron Collider (LHC) with its unprecedented collision
energies have brought up new directions in understanding the possible formation of a QGP medium even in pp collisions. In this direction, the
observations of QGP-like properties such as strangeness enhancement~\cite{ALICE:2017jyt}, double-ridge structure~\cite{Khachatryan:2016txc} etc. in smaller collision systems like pp and p--Pb collisions are note worthy. These developments have important consequences on the results obtained from heavy-ion collisions, as pp collisions are used as baseline measurements while characterizing the medium properties in heavy-ion
collisions. A closer look at the LHC pp collisions, especially the high-multiplicity events is a call of time.
\begin{figure}[ht!]
\begin{center}
\includegraphics[scale = 0.5]{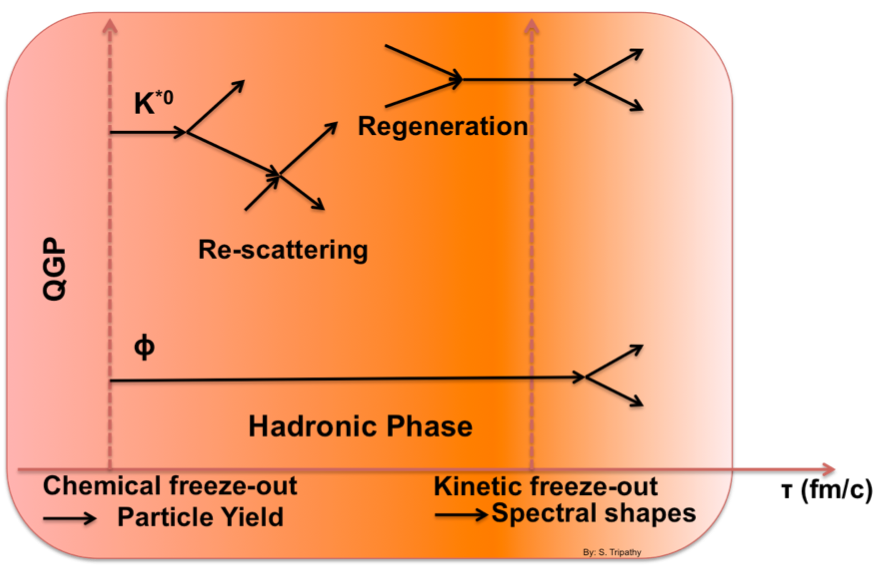}
\caption{Depiction of re-scattering and regeneration processes of the resonances in hadronic phase in heavy-ion collisions.}
\label{fig1}
\end{center}
\end{figure}

There are very few reports on the estimation of hadronic lifetime in pp and heavy-ion collisions from either theoretical or experimental prospectives \cite{Ralf}. In this work, we have made an attempt to use short-lived hadronic resonances (like $\rm K^{*0}$) produced in these collisions to have an estimation of hadronic
phase lifetime. In addition, in view of the observed multiplicity scaling at the LHC, we have studied the hadronic phase lifetime as a function of event multiplicity across various collision species like pp, p--Pb, Cu--Cu, Au--Au and Pb--Pb collisions for collision energies spanning from GeV to TeV. On the contrary, we use long-lived hadronic resonances like $\phi$ mesons, which have lower hadronic interaction cross section, to locate the QGP phase boundary in terms of the effective temperature obtained from the Boltzmann-Gibbs Blast-Wave distribution function. This work would shed light on the role of charged-particle multiplicity, collision system and collision energy dependence of the partonic and hadronic phases produced at the RHIC and LHC energies.

The paper is organized as follows. The detailed methodology of estimating lifetime of hadronic phase and location of QGP phase boundary is discussed in the next section. Section~\ref{results} reports the results along with their discussions. Finally, we summarize in section~\ref{sum} with important findings.

\section{Methodology}
Before going through the entire procedure of estimation of lifetime, let us begin with a brief introduction on evolution of the ultra-relativistic heavy-ion collisions. When two Lorentz contracted nuclei collide at very high energies, the region where they overlap is very thin in the longitudinal direction, much like an almond shape. This energetic interaction results in the formation of a possible state of QGP. QGP exists at very high temperature and/or energy density and consists of asymptotically free quarks and gluons, otherwise known as partons. The created fireball then expands because of very high energy deposition in a small volume with ultra-high temperature ($\sim 10^5$ times the core of the Sun) and then it cools down till the particles reach a kinetic freeze-out. One can approximately estimate the formation time of the hadrons by the use of the uncertainty principle. The formation time in the rest frame can be related to the hadron size, $R_h$. In laboratory frame, the hadron formation time is then given by $t_{form} \simeq R_h\frac{E_h}{m_h}$, where $E_h$ and $m_h$ being the energy and mass of the hadron, respectively~\cite{Adcox:2004mh}. Then, the hadrons start forming inside this QGP medium. After a certain temperature known as the chemical freeze-out temperature ($T_{\rm ch}$), the hadron formation ceases and the stable particle numbers are fixed. At this point, the hadronic phase begins where the produced short-lived resonances decay and the daughter particles undergo multiple re-scatterings. After some time, the momentum transfer between the particles also ceases at a temperature called as the kinetic freeze-out temperature ($T_{\rm th}$). Finally all the particles move with relativistic velocities towards the detectors. However, the calculation of QGP and hadronic phase lifetime is not trivial as we can only have the information about the final state particles in experiments. On the other hand, resonance particles can be used as a probe to calculate the hadronic phase lifetime and locate the QGP phase boundary.

Resonances are usually referred as the particles, which have higher masses than that of the corresponding ground state particle(s) with similar quark content. As hadronic resonances decay strongly, they have very short lifetime, $\tau \sim$ few fm/c. Before decaying, these particles can only travel upto few femto-meters. The width ($\Gamma$) and lifetime ($\tau$) of the resonances are related by the Heisenberg's uncertainty relation, i.e. $\Gamma\tau = \hbar$. As the broad resonance states decay very shortly after their production, it can only be measured by reconstruction of their decay daughters in a detector. The typical lifetimes of experimentally measured hadronic resonances range from 1.1 fm/c to 46.2 fm/c. Hadronic resonances are produced in the bulk of the expanding medium in heavy-ion collisions and they can decay while still traversing in the medium. The decay daughters may interact with other particles in the medium, which would result in suppression of resonances in their reconstruction, as the invariant mass of the daughters may not match with that of the parent particle. This process is known as re-scattering. In other way, resonances can be regenerated as a consequence of pseudo-elastic collisions in the hadronic phase of the medium, which would result in enhancement of the resonance yields. Resonances, with relatively higher life-time, might not go through any of the above mentioned processes. It may also happen that the re-scattering and re-generation processes compensate each other. Thus, the interplay of these processes makes the study of resonances in heavy-ion collisions (Fig.~\ref{fig1}) more fascinating. As discussed in Ref. \cite{Ralf1}, the suppression in the ratio of resonances like 
$\rm{K}^{*0}/K$ and $\rho/\pi$ could be due to their late production closer to the kinetic freeze-out.

In Fig.~\ref{fig1}, we show the hadronic phase which starts from chemical freeze-out, where the long-lived particle yields are fixed; to the kinetic freeze-out, where the final state particle spectra get fixed. We schematically show the re-scattering and regeneration processes which might be possible for short-lived resonances like $\rm{K}^{*0}$, while it is expected that the long-lived resonances like $\phi$ meson would not go through any of such processes. It is established that the ratio of $\rm{K}^{*0}$ to $\rm{K}$ shows significant suppression for central heavy-ion collisions with respect to pp collisions, while $\phi$ does not go through any such enhancement and/or suppression~\cite{Abelev:2014uua}. This indicates the dominance of re-scattering processes over re-generation process in hadronic phase. Depending on the suppression of $\rm{K}^{*0}$, one can calculate hadronic phase lifetime. For pp low multiplicity collisions, the $\rm{K}^{*0}$/K ratio is taken as the ratio at the chemical freeze-out temperature. The $\rm{K}^{*0}$/K ratio at different centralities for different collision systems can be taken as the ratio at the kinetic freeze-out temperature. The lifetime can be calculated using the following relation~\cite{Singha:2015fia,Adams:2004ep,Acharya:2019qge}, 
\begin{equation}
[\rm{K^{*0}}/\rm{K}]_{kinetic} = [\rm{K^{*0}}/\rm{K}]_{chemical} \times e^{-\Delta t/\tau},
\label{eq1}
\end{equation}
where $\tau$ is the lifetime of $\rm{K}^{*0}$ and $\Delta t$ is the hadronic phase lifetime multiplied by the Lorentz factor. The Lorentz factor is calculated using the mean transverse momentum ($\langle p_{\rm{T}}\rangle$) of $\rm{K}^{*0}$.
One could naively expect that the interaction cross-section of the decay daughters of $\rm{K}^{*0}$ ($\pi$ and $\rm{K}$) with other pions in the hadronic phase would be much higher than that of with kaons due to the large relative abundance of pions than kaons. For regeneration, the interaction of $\pi \rm{K}$ is essential. So, one would expect that the regeneration effects would be very small compared to re-scattering effects. Hence, in our calculation we have neglected the effect of regeneration. Our assumption is supported by the calculations of pion-pion and pion-kaon interactions, where the former interaction cross-section is nearly 5 times larger than that of the latter~\cite{Protopopescu:1973sh,Matison:1974sm}. If there is considerable amount of regeneration of $\rm{K}^{*0}$, then the lifetime of hadronic phase would be higher than the calculated lifetime using Eq.~\ref{eq1}. Hence, we call the calculated lifetime as the {\it lower limit} on the hadronic phase lifetime. As the kinetic freeze-out is the time, when all the elastic collisions happen to cease (the mean free path of the system becomes higher than the system size), in a high-multiplicity scenario, because of higher interaction rates due to lower mean free path (particle density being higher), the lifetime of the hadronic phase is expected to be higher as compared to low-multiplicity events. This also justifies the hadronic phase lifetime of heavy-ion collisions being higher than high-multiplicity pp collisions and the latter being again higher than the low-multiplicity events.

Contrary to $\rm{K}^{*0}$, $\phi$ can act as a perfect tool to probe the location of QGP phase boundary. As finding the QGP lifetime is not trivial, we have obtained the temperature ($T_{\rm eff}$) of the system in the QGP phase boundary using the $\phi$ meson. The $\phi$ meson is the lightest bound state of a strange and anti-strange quark (s$\bar{s}$), and is produced early in the reaction relative to stable particles such as pions, kaons and protons. It is least affected by hadronic re-scattering or regeneration because of its relatively longer lifetime and the decay daughters are expected to decay outside the fireball~\cite{Sahoo:2010gr,Ko:1993id,Haglin:1994xu}. The results extracted from the analysis of the experimental data on the ratio of various hadronic species~\cite{BraunMunzinger:2003zd} indicate that the inelastic interactions of $\phi$ meson with other hadrons is not significant below the $T_{\rm ch}$. In Refs.~\cite{Shor:1984ui,Sibirtsev:2006yk}, it has been shown that the $\phi N$ cross-section is only around 8-12 mb for photoproduction on proton and nuclear targets below 10 GeV. It has also been shown that out of all $\phi$ mesons produced, only 5\% would re-scatter in the hadronic phase of the medium. This in principle, should go down as one moves to TeV energies, where the matter is almost baryon free ($\bar{p}/p \sim 0.9$) \cite{Aamodt:2010dx,Tawfik:2010pt}. As the interactions of $\phi$ meson with other hadrons are very less in the mixed and hadron gas phase, the information of the QGP phase boundary remains intact when $\phi$ meson is used as a probe. Hence, the transverse momentum spectra of $\phi$ meson after its hadronisation will not be altered or distorted during the expanding hadronic phase. Using the transverse momentum spectra of $\phi$, one can obtain reliable information on intensive variables, such as the critical temperature and the location of QGP phase boundary~\cite{Shor:1984ui,Nasim:2015gua}. We fit the Boltzmann-Gibbs Blast-Wave (BGBW) function to the $p_{\rm{T}}$ spectra of $\phi$ meson up to $p_{\rm T} \sim$ 3 GeV/c to get the $T_{\rm th}$, the true freeze-out temperature and $\langle\beta\rangle$, the average velocity of medium, which can then be used to find out the effective temperature, $T_{\rm eff}$ using the following relation.
\begin{equation}
T_{\rm{eff}} = T_{\rm{th}} + \frac{1}{2}m\langle \beta \rangle^{2}
\label{eq3}
\end{equation}
The expression for invariant yield in the BGBW framework is given as follows~\cite{Schnedermann:1993ws}:
 \ba
\label{bgbw1}
E\frac{d^3N}{dp^3}=D \int d^3\sigma_\mu p^\mu exp(-\frac{p^\mu u_\mu}{T}).
\ea
Here, the particle four-momentum is, 
\ba
p^\mu~=~(m_T{\cosh}y,~p_T\cos\phi,~ p_T\sin\phi,~ m_T{\sinh}y).
\ea
The four-velocity is given as,
\ba
u^\mu=\cosh\rho~(\cosh\eta,~\tanh\rho~\cos\phi_r,~\tanh\rho~\sin~\nonumber\\
\phi_r,~\sinh~\eta).
\ea 
The kinetic freeze-out surface is parametrised as, 
\ba
d^3\sigma_\mu~=~(\cosh\eta,~0,~0, -\sinh\eta)~\tau~r~dr~d\eta~d\phi_r,
\ea
where, $\eta$ is the space-time rapidity. Assuming Bjorken correlation in rapidity for simplification, $i.e.$ $y=\eta$~\cite{Bjorken:1982qr}, Eq.~\ref{bgbw1} is expressed as,  
\ba
\label{boltz_blast}
\left.\frac{d^2N}{dp_Tdy}\right|_{y=0} = D \int_0^{R_{0}} r\;dr\;K_1\Big(\frac{m_T\;\cosh\rho}{T_{\rm{th}}}\Big)I_0\nonumber\\
\Big(\frac{p_T\;\sinh\rho}{T_{\rm{th}}}\Big).
\ea
Here, $D$ is the normalisation constant, $g$ is the degeneracy factor and $m_{\rm T}=\sqrt{p_T^2+m^2}$ is the transverse mass. $K_{1}\displaystyle\Big(\frac{m_T\;{\cosh}\rho}{T_{\rm{th}}}\Big)$ and $I_0\displaystyle\Big(\frac{p_T\;{\sinh}\rho}{T_{\rm{th}}}\Big)$ are the modified Bessel's functions. They are given by,
\ba
\centering
K_1\Big(\frac{m_T\;{\cosh}\rho}{T}\Big)=\int_0^{\infty} {\cosh}y\;{\exp} \Big(-\frac{m_T\;{\cosh}y\;{\cosh}\rho}{T_{\rm{th}}}\Big)dy\nonumber,
\ea
\ba
\centering
I_0\Big(\frac{p_T\;{\sinh}\rho}{T}\Big)=\frac{1}{2\pi}\int_0^{2\pi} exp\Big(\frac{p_T\;{\sinh}\rho\;{\cos}\phi}{T_{\rm{th}}}\Big)d\phi \nonumber.
\ea
Here, $\rho$ in the integrand is a parameter given by $\rho={\tanh}^{-1}\beta$ and $\beta=\displaystyle\beta_s\;\Big(\xi\Big)^n$ \cite{Huovinen:2001cy,Schnedermann:1993ws,BraunMunzinger:1994xr, Tang:2011xq} is the radial flow. Here, $\beta_s$ is the maximum surface velocity and $\xi=\displaystyle\Big(r/R_0\Big)$, where $r$ is the radial distance. In the BGBW model the particles closer to the center of the fireball move slower than the ones at the edges and the average of the transverse velocity is evaluated as \cite{Adcox:2003nr}, 
\ba
<\beta> =\frac{\int \beta_s\xi^n\xi\;d\xi}{\int \xi\;d\xi}=\Big(\frac{2}{2+n}\Big)\beta_s.
\ea
For our calculation, we use a linear velocity profile, ($n=1$) and $R_0$ is the maximum radius of the expanding source at freeze-out ($0<\xi<1$).


Keeping the above procedure in mind, let us discuss the results on how final state multiplicity plays a role in the QGP and hadronic phase lifetime. 

\begin{figure}[H]
\includegraphics[scale=0.45]{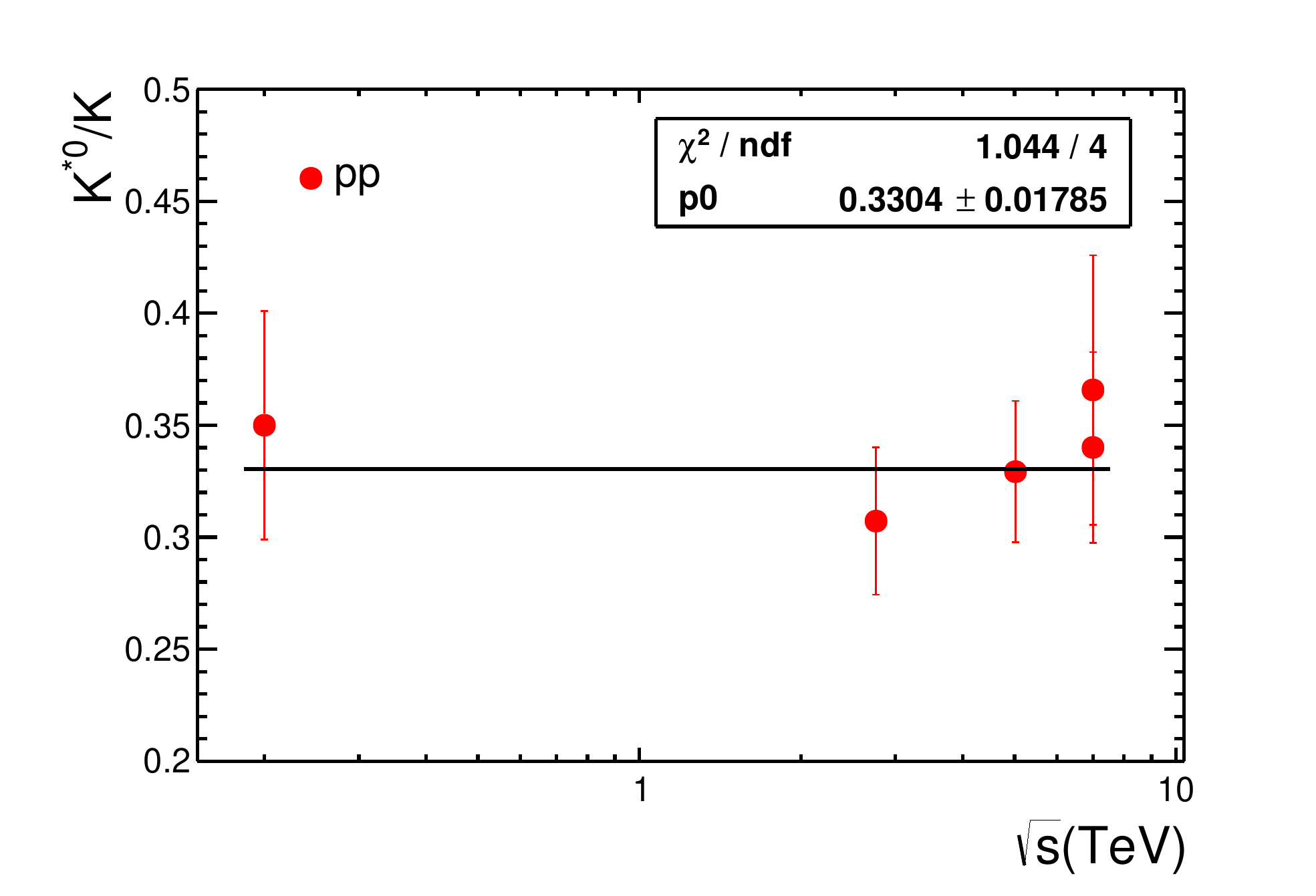}
\caption[]{(Color Online) $\rm{K}^{*0}$ to K ratio for pp collisions as a function of collision energy. The solid black line shows the fitting of the points with polynomial of order zero.}
\label{fig2.0}
\end{figure}

\begin{figure}[H]
\includegraphics[scale=0.45]{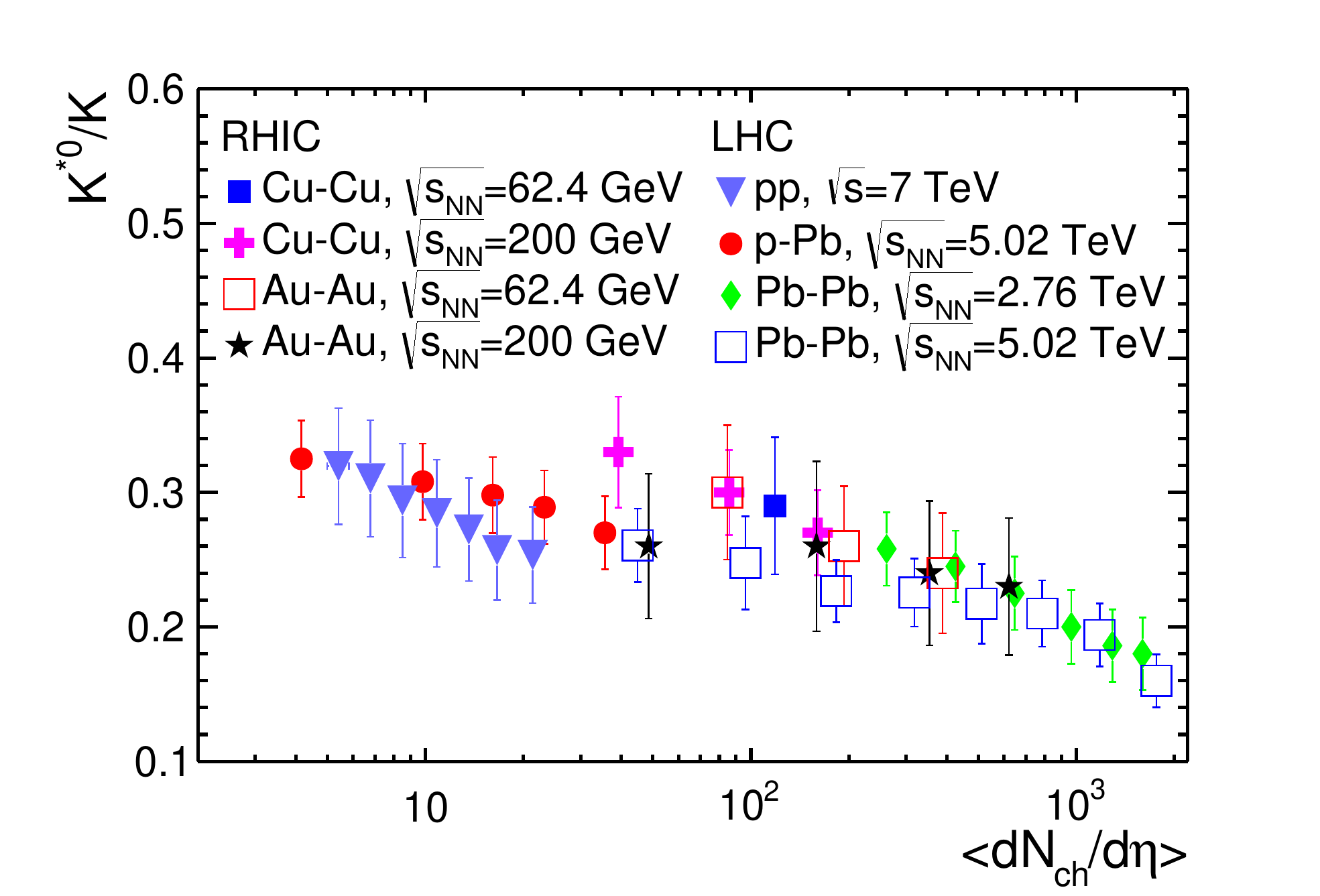}
\caption[]{(Color Online) $\rm{K}^{*0}$ to K ratio for different collision systems as a function of charged-particle multiplicity.}
\label{fig2}
\end{figure}

\section{Results and Discussion}
\label{results}

\begin{figure}[ht!]
\includegraphics[scale=0.45]{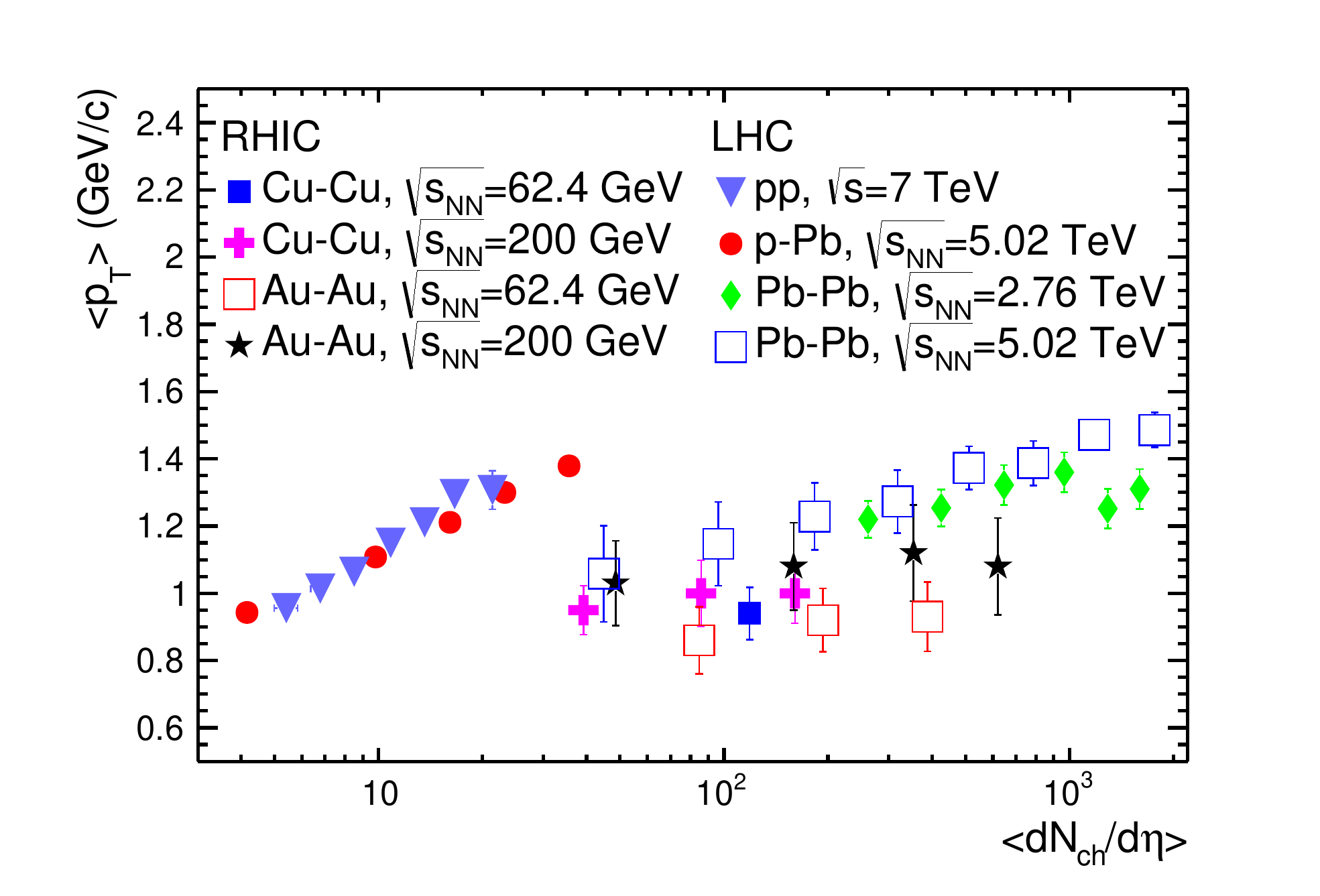}
\caption[]{(Color Online) Mean transverse momentum of $\rm{K}^{*0}$ for different collision systems as a function of charged-particle multiplicity at different collision energies.}
\label{fig3}
\end{figure}

\begin{figure}[ht!]
\includegraphics[scale=0.45]{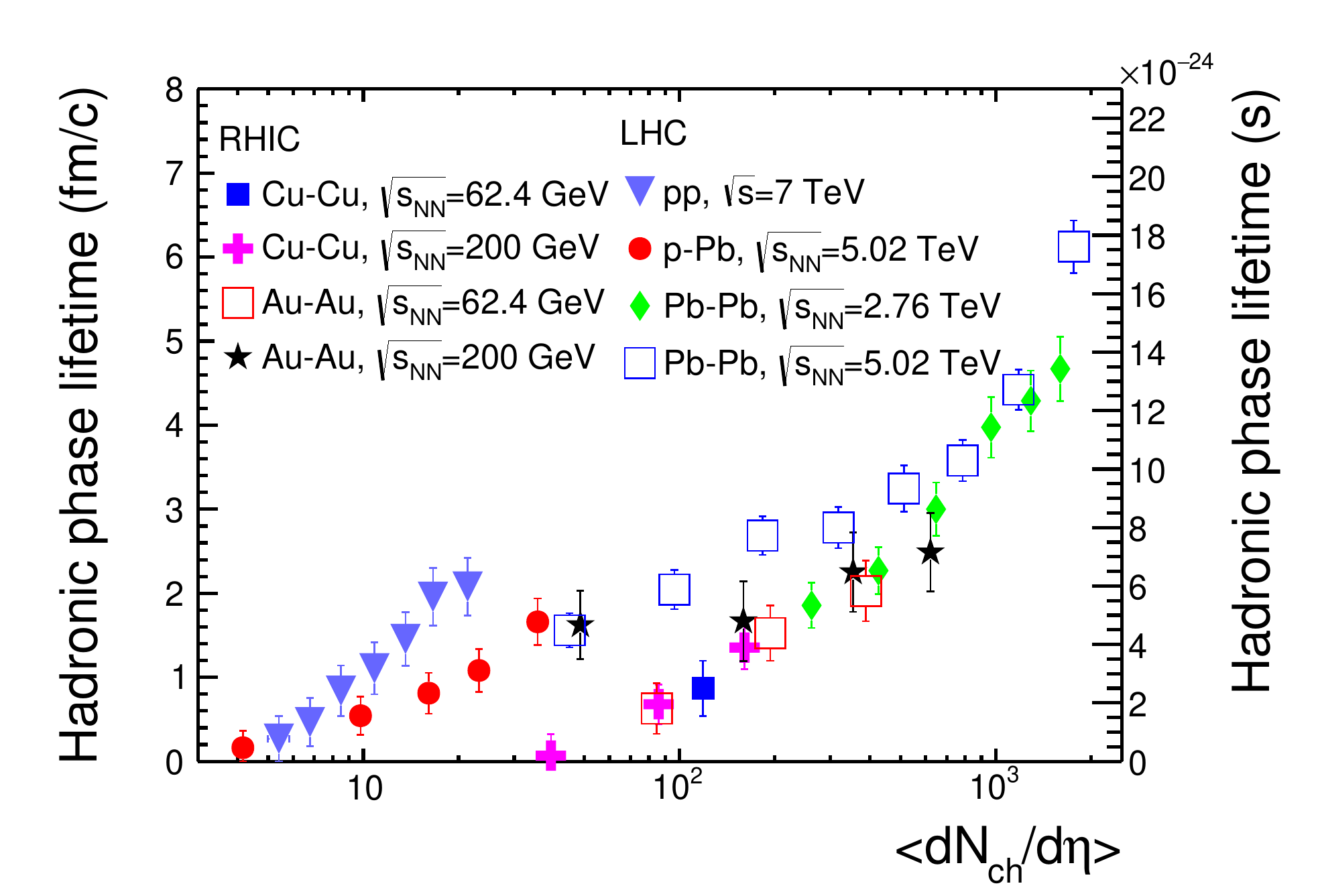}
\caption[]{(Color Online) Hadronic phase lifetime as a function of charged-particle multiplicity for different collision systems at RHIC and LHC energies.}
\label{fig4}
\end{figure}

\begin{figure}[ht!]
\includegraphics[scale=0.45]{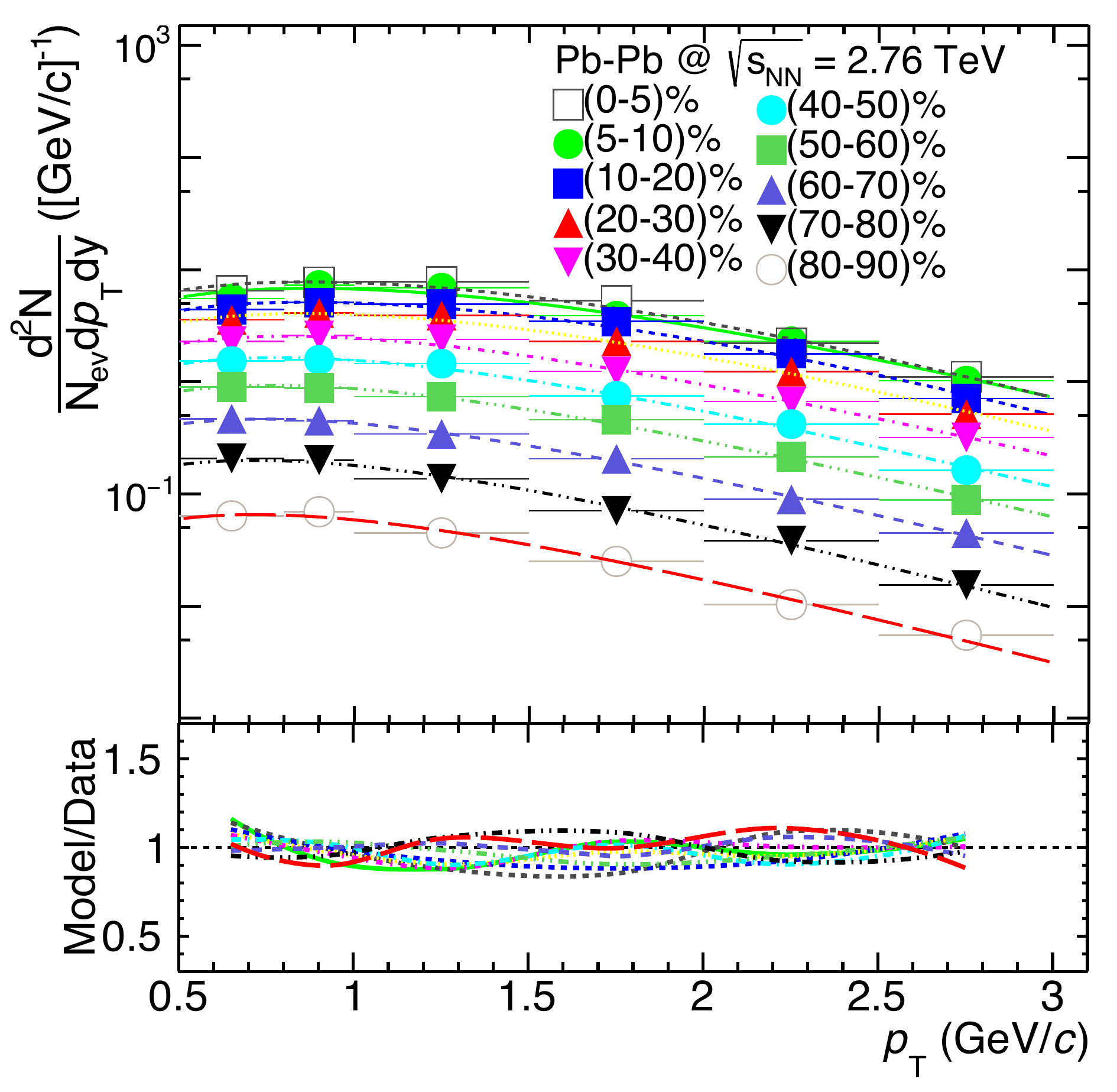}
\caption[]{(Color Online) Fitting of Boltzmann-Gibbs Blast-Wave distribution to $p_{\rm{T}}$ spectra of $\phi$ meson production in Pb-Pb collisions at $\sqrt{s_{NN}}$ = 2.76 TeV.}
\label{fig9}
\end{figure}
 \begin{figure}[ht!]
 \includegraphics[scale=0.45]{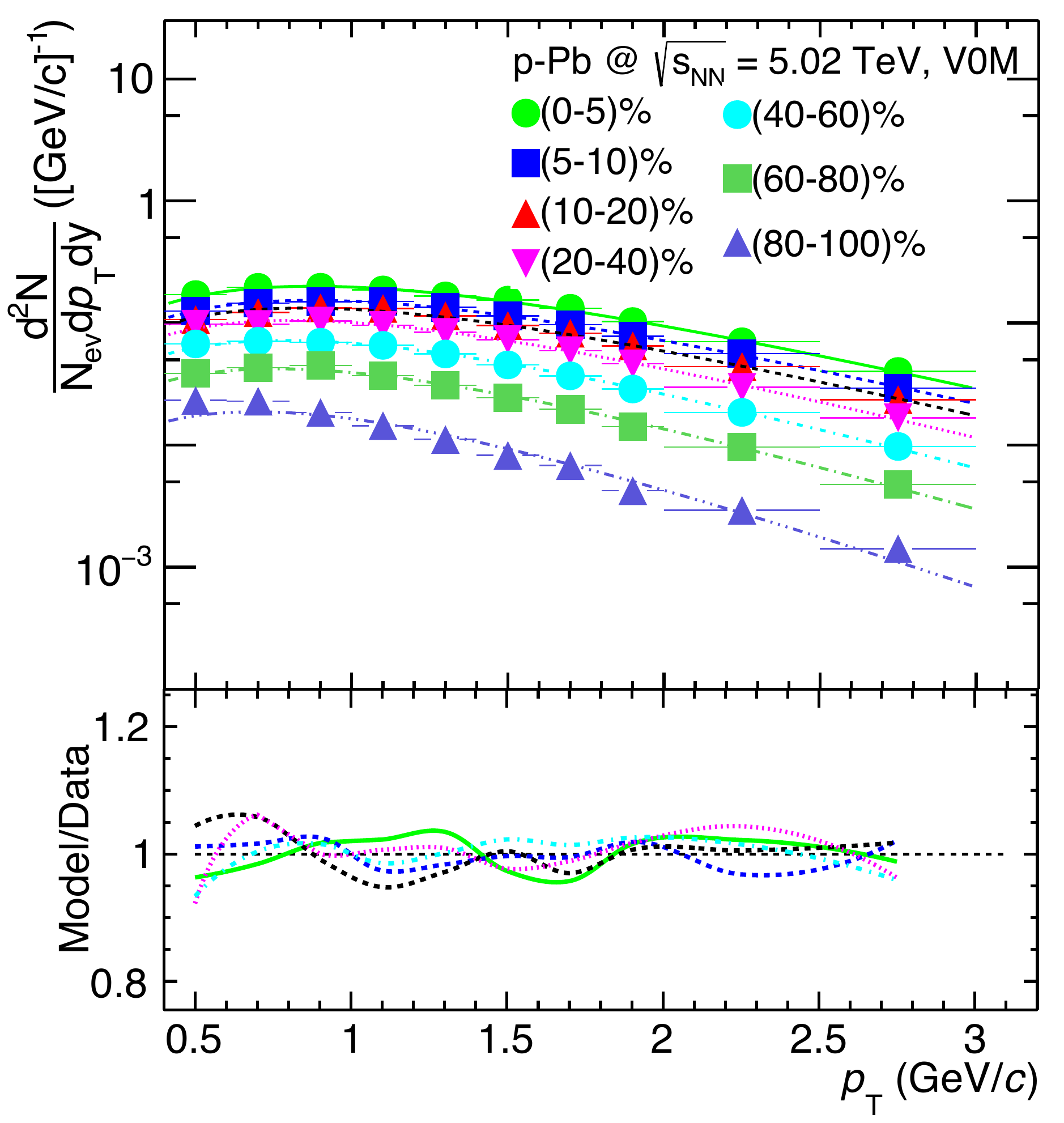}
\caption[]{(Color Online) Fitting of Boltzmann-Gibbs Blast-Wave distribution to $p_{\rm{T}}$ spectra of $\phi$ meson production in p-Pb collisions at $\sqrt{s_{NN}}$ = 5.02 TeV.}
\label{fig10}
\end{figure}
 \begin{figure}[ht!]
\includegraphics[scale=0.45]{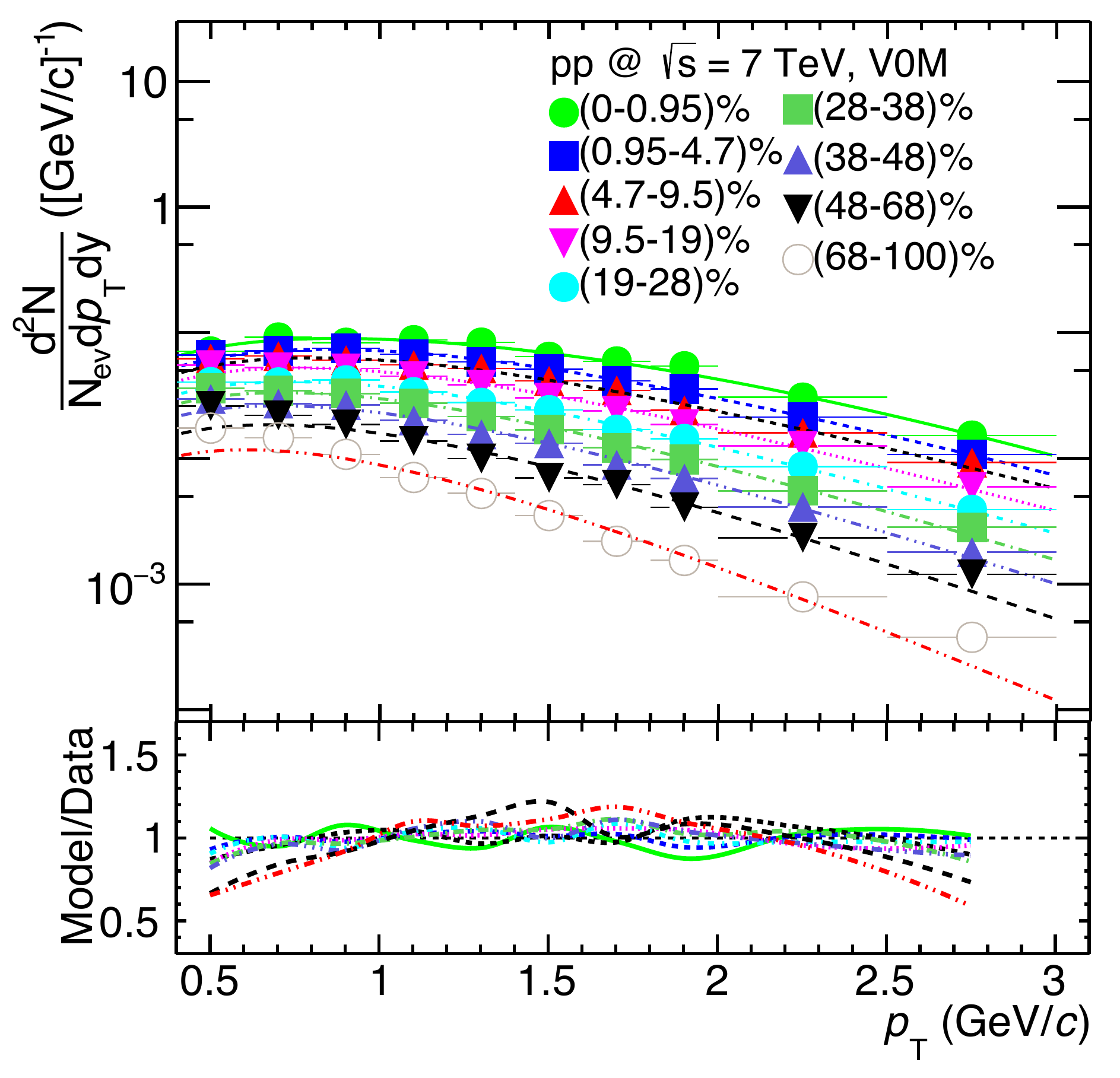}
\caption[]{(Color Online) Fitting of Boltzmann-Gibbs Blast-Wave distribution to $p_{\rm{T}}$ spectra of $\phi$ meson production in pp collisions at $\sqrt{s}$ = 7 TeV.}
\label{fig11}
\end{figure}


Figure~\ref{fig2.0} shows the $\rm{K}^{*0}$ to K ratio for pp collisions as a function of collision energy for RHIC and LHC. 
Here we have taken the $\rm{K}^{*0}$/K ratios for all the minimum bias pp collisions for different collision energies. In addition, we have also included $\sqrt{s}=$ 7 TeV ratio for the low multiplicity pp collisions. The solid black line shows the fitting of the data points with a zeroth order polynomial. Assuming this ratio to be the
same across all the collision energies, we use $\rm{K}^{*0}$/K = 0.33 $\pm$ 0.02 as the ratio at the chemical freeze-out in Eq.~\ref{eq1}, which is obtained from the above fitting. Figure~\ref{fig2} shows the $\rm{K}^{*0}$/K ratio as a function of charged-particle multiplicity for different collision systems at RHIC and LHC~\cite{Adams:2004ep,Acharya:2019qge,Acharya:2018orn,Adam:2017zbf,Adam:2016bpr,Tripathy:2018ehz,Alver:2010ck,Aggarwal:2010mt}. As the data of charged-particle multiplicity, d$N_{ch}$/d$\eta$ for each centrality class are not available for RHIC energies, we have used Eq.7 of Ref.~\cite{Alver:2010ck} (obtained from simultaneous fits) for the conversion from the average number of participants. One should note here that the rapidity range of d$N_{ch}$/d$\eta$ is $|\eta| < 0.5$ for LHC energies while $|\eta| < 1.0$ for RHIC energies. Clearly, the ratio decreases as a function of charged-particle multiplicity, which suggests significant dominance of re-scattering effects of the decay daughters over regeneration effects in the hadronic phase. This behavior enables us to use Eq.~\ref{eq1} to obtain the lower limit of the hadronic phase lifetime. For the calculation of the Lorentz factor for hadronic phase lifetime, one needs the mean-transverse momentum ($\langle p_{\rm{T}}\rangle$). Fig.~\ref{fig3} shows $\langle p_{\rm{T}}\rangle$ as a function of charged-particle multiplicity for different collision systems at RHIC and LHC~\cite{Adams:2004ep,Acharya:2019qge,Acharya:2018orn,Adam:2017zbf,Adam:2016bpr,Alver:2010ck,Aggarwal:2010mt}. Clearly, the evolution of $\langle p_{\rm{T}}\rangle$ as a function of charged-particle multiplicity does not show smooth evolution across collision systems. The $\langle p_{\rm{T}}\rangle$ for small systems like pp and p-Pb collisions have completely different trend than that of heavy-ion collisions. When one moves to higher domain of collision energies, one produces more particles and also the $\langle p_{\rm{T}}\rangle$ of the system increases. There seems to be a correlated increase of particle density in phase space and $\langle p_{\rm{T}}\rangle$. With same particle density, a higher $\langle p_{\rm{T}}\rangle$ would indicate higher collision rate (re-scattering) and hence higher hadronic phase lifetime.
The calculated hadronic phase lifetime using Eq.~\ref{eq1} shows a linear increase as a function of charged particle multiplicity, as depicted in Fig.~\ref{fig4}. This suggests that for a given charged-particle multiplicity, the hadronic phase lifetime is similar irrespective of the collision energy and collision systems for central heavy-ion collisions like Cu-Cu and Au-Au collisions at RHIC and Pb-Pb collisions at the LHC. Peripheral heavy-ion collisions seem to show a different trend, which might be due to the effect of system size and collisions energy or in other words, the effective energy deposited in the Lorentz contracted region \cite{Mishra:2014dta,Sarkisyan:2015gca,Sarkisyan:2016dzo,Sarkisyan-Grinbaum:2018yld}. However, the small collision systems like pp and p-Pb collisions at the LHC show different evolution compared to heavy-ion collisions. This behavior seems to be propagated from the dependence of $\langle p_{\rm{T}}\rangle$ as a function of charged-particle multiplicity although the $\rm{K}^{*0}$/K ratio shows a smooth decrease as a function of charged-particle multiplicity. The strong evolution of the lifetime across collision systems and collision energies is clearly visible in the figure. It is observed that the hadronic phase lifetime in high-multiplicity pp collisions is of the order of 2 fm/c, whereas for central Pb--Pb collisions it is around 6 fm/c. 

\begin{figure}[ht!]
\includegraphics[scale=0.45]{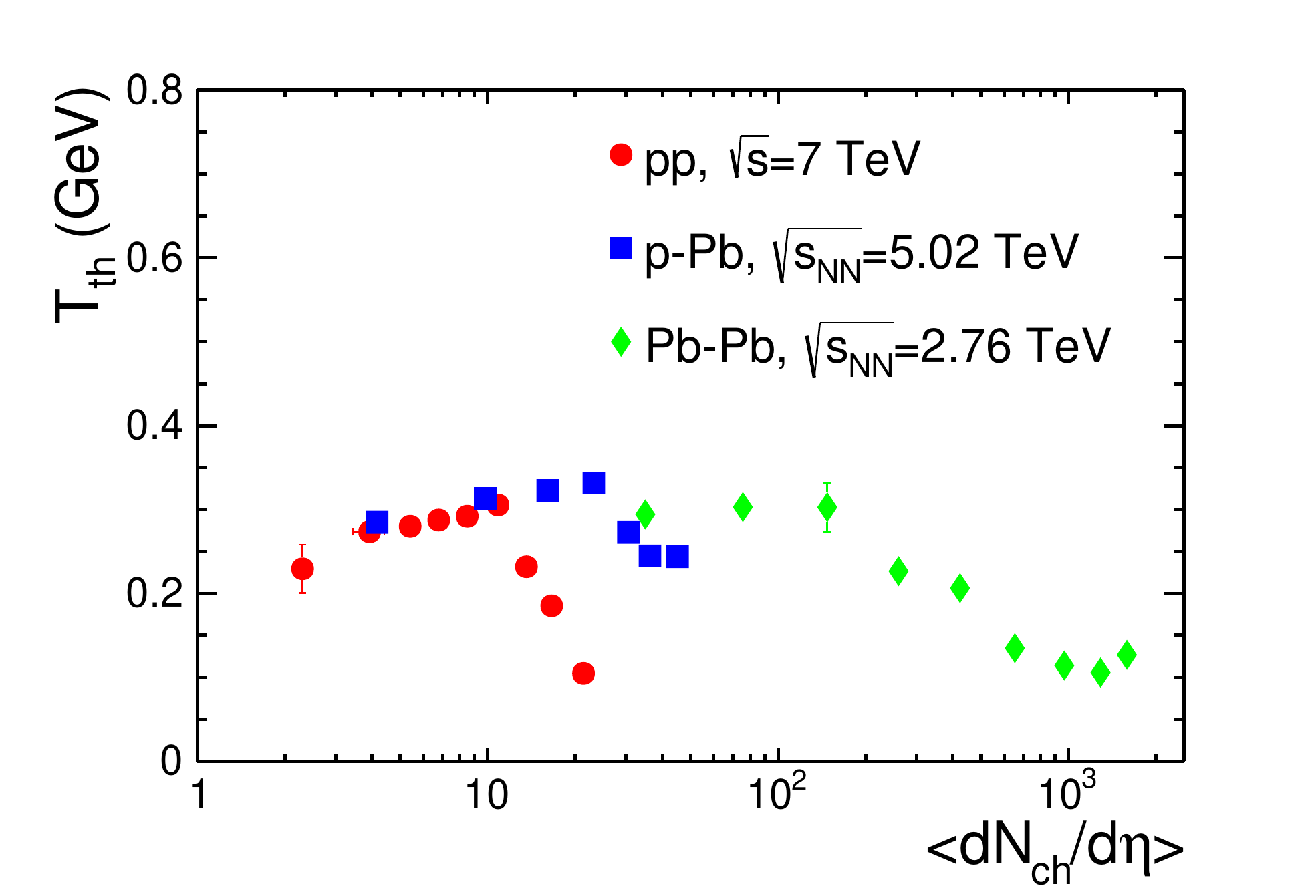}
\caption[]{(Color Online) Kinetic freeze-out temperature for $\phi$ meson as a function charged-particle multiplicity for different collision systems.}
\label{fig6}
\end{figure}

We fit the BGBW as given in Eq.~\ref{boltz_blast} to the $p_{\rm{T}}$~spectra of $\phi$ meson in different multiplicity classes to get the $T_{\rm{th}}$, the true freeze-out temperature and $\langle \beta \rangle$, the average radial flow velocity of medium. Fig.~\ref{fig9} shows the Blast-Wave fit for $p_{\rm{T}}$~spectra in Pb-Pb collisions at $\sqrt{s_{NN}}$ = 2.76 TeV for different centrality classes. Similarly, Fig.~\ref{fig10} and ~\ref{fig11} show the Blast-Wave fit for $p_{\rm{T}}$~spectra in p-Pb collisions at $\sqrt{s_{NN}}$ = 5.02 TeV and in pp collisions at $\sqrt{s}$ = 7 TeV for different multiplicity classes, respectively. The fitting is performed upto $p_{\rm T}$ = 3 GeV/c. The lower panels show the fit to data ratio for different collision systems. The fit quality seems reasonable as the maximum deviation of fit does not exceed beyond 10\% for Pb--Pb, p--Pb and high multiplicity pp collisions. As expected, the fit quality goes bad for the low multiplicity pp collisions due to less probable Blast-Wave scenario in low multiplicity pp collisions.

\begin{figure}[ht!]
\includegraphics[scale=0.45]{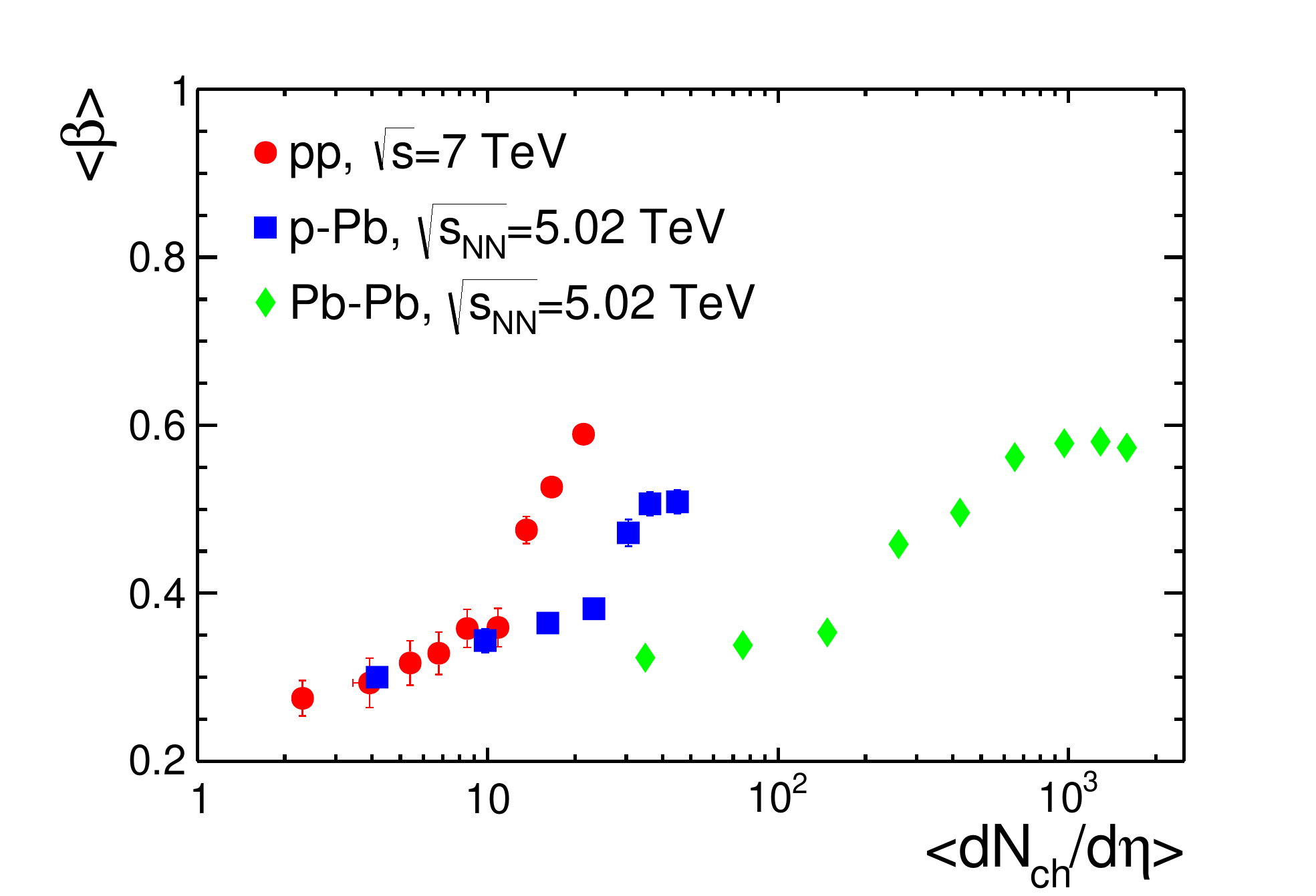}
\caption[]{(Color Online) Average radial flow for $\phi$ meson as a function charged-particle multiplicity for different collision systems.}
\label{fig7}
\end{figure}

\begin{figure}[ht!]
\includegraphics[scale=0.45]{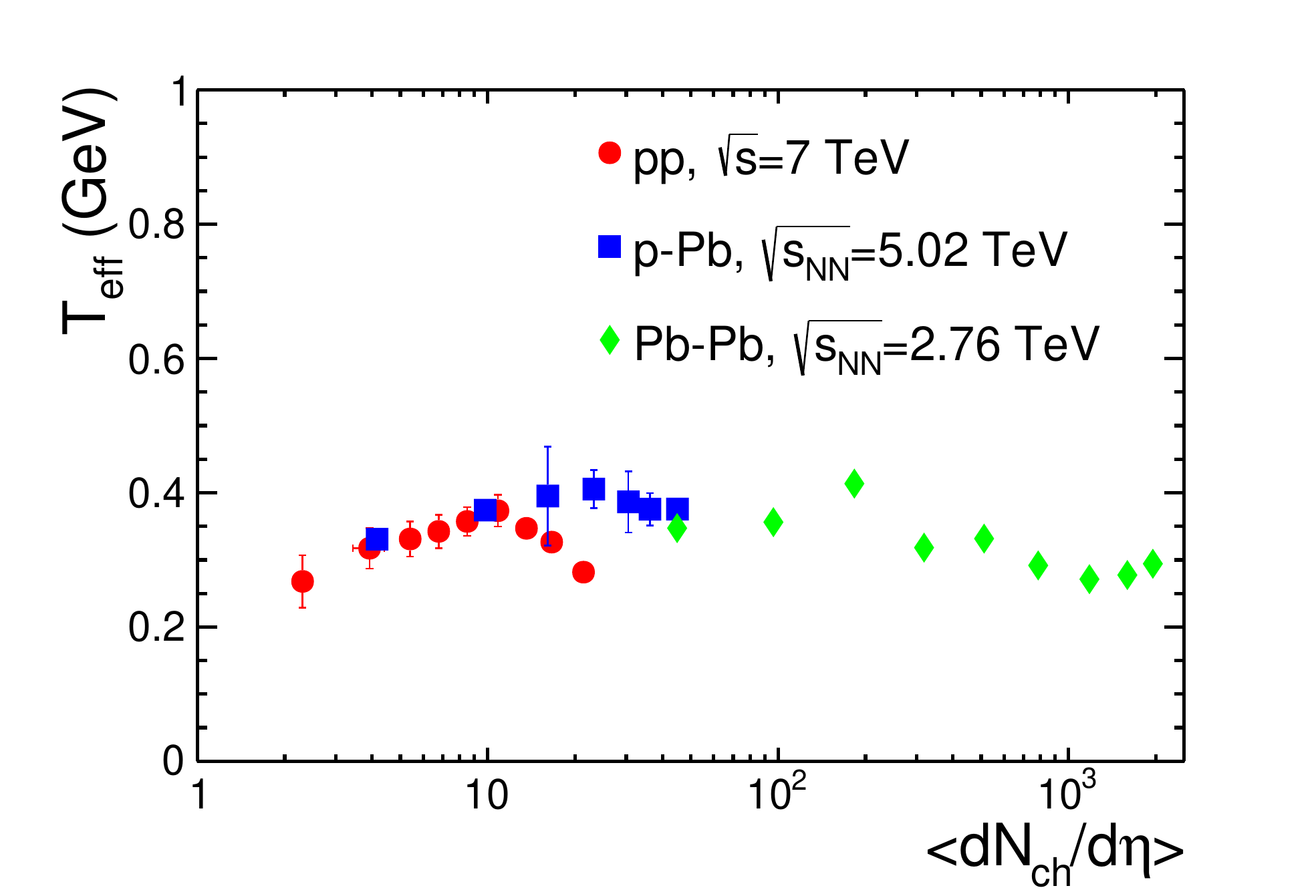}
\caption[]{(Color Online) Effective temperature for $\phi$ meson as a function charged-particle multiplicity for different collision systems.}
\label{fig8}
\end{figure}

From these fits, we find the values of $\langle \beta \rangle$ and $T_{\rm{th}}$ for all collision systems in different centralities. By using these values in Eq.~\ref{eq3}, we can find the effective temperature ($T_{\rm{eff}}$) of $\phi$ mesons. Fig.~\ref{fig6} shows the kinetic freeze-out temperature for $\phi$ meson, denoted by $T_{\rm{th}}$, as a function of charged-particle multiplicity. One can observe that the temperature shows almost a flat trend upto a certain d$N_{ch}$/d$\eta$ and then drops immediately afterwards. This can be explained by considering the fact that for low charged-particle multiplicity, the system freezes out early, means it freezes out at high $T_{\rm{th}}$. As the charged-particle multiplicity becomes more, the system is thought to have gone through QGP phase, which results in the system taking a longer time to attain the kinetic freeze-out. This is also evident from our findings of higher hadronic phase lifetime for high-multiplicity collisions. As a result, the kinetic freeze-out temperature drops abruptly in all the collision systems. We observe that the drop of $T_{\rm{th}}$ happens at different charged-particle multiplicity in different collision systems. Fig.~\ref{fig7} shows the average radial flow as a function of charged-particle multiplicity for different collision systems at the LHC. It can be seen that $\langle \beta \rangle$ increases smoothly in all the collision systems upto a certain extent. However, for pp collisions at a certain charged-particle multiplicity ($\simeq10-20$), $\langle \beta \rangle$ shows a sudden increase. Observation of this threshold in the final state charged particle multiplicity is supported by the following additional observations for a change in dynamics of the system. $N_{\rm ch}\simeq10-20$ has been found to be a limit after which the multipartonic interactions (MPI) in pp collisions is found to show an important role in particle production (quarkonia) at the LHC energies \cite{Thakur:2017kpv}. In addition, this threshold is also supported as a thermodynamic limit after which all the statistical ensembles give similar results  while describing the freeze-out properties of the system \cite{Sharma:2018jqf}.  After this threshold, we have also observed $T_{\rm ch}$ to be higher than the kinetic freeze-out temperature \cite{Rath:2019aas}.
We know that the average radial flow is larger for a system that goes through QGP phase. Again, this suggests a higher probability of QGP formation after this particular charged-particle multiplicity which may have been the reason for the sudden increase in $\langle \beta \rangle$. The behavior of $\langle \beta \rangle$ compliments the results observed in Fig.~\ref{fig6}. This again is supported by earlier predictions for a cross-over transition from hadronic to QGP phase to happen between charged particle multiplicity density of 6 to 24 \cite{Campanini}. 

Figure~\ref{fig8} shows the effective temperature of $\phi$ meson as a function of charged-particle multiplicity, which encodes the temperature due to thermal motion ($T_{\rm{th}}$) and due to the collective motion (calculated from $\langle \beta \rangle$). It is clearly seen that regardless of the collision systems, $T_{\rm{eff}}$ does not show any major dependence on d$N_{ch}$/d$\eta$. This behavior is unlike the behaviours observed for $T_{\rm{th}}$ and $\langle \beta \rangle$ as a function of charged-particle multiplicity. As $\phi$ meson keeps the information of QGP phase boundary intact, the trend of $T_{\rm{eff}}$ suggests that the location of QGP phase boundary is independent or weakly dependent on charged particle multiplicity.
Interestingly, this observation is also supported by earlier reports of $T_{\rm ch}$ being independent of final state charged particle multiplicity \cite{Rath:2019aas}.

Figure~\ref{fig4} and Fig.~\ref{fig8} suggest that the hadronic phase lifetime strongly depends on charged-particle multiplicity while the QGP phase shows very weak dependence on charged-particle multiplicity.  
This indicates that the hadronisation from a QGP state starts at a similar temperature irrespective of charged-particle multiplicity, collision system and collision energy while the duration of hadronic phase is strongly depends on final state charge-particle multiplicity.
 
 \section{Summary}
  \label{sum}
  
In this work, we have made an attempt to use hadronic resonances produced in pp, p--Pb, Cu--Cu, Au--Au and Pb--Pb collisions to have an estimation of hadronic
phase lifetime and to locate the QGP phase boundary. In summary,
\begin{enumerate}

\item For a given charged-particle multiplicity the hadronic phase lifetime is similar irrespective of the collision energy and collision systems for central heavy-ion collisions like Cu--Cu and Au--Au collisions at RHIC and Pb--Pb collisions at the LHC.  Peripheral heavy-ion collisions seem to show a different behaviour, which could be because of the effect of system size and collision energy or in brief, the effective energy responsible for particle production. However, the small collision systems like pp and p--Pb collisions at the LHC show different evolution compared to heavy-ion collisions. This behavior seems to be propagated from the dependence of $\langle p_{\rm{T}}\rangle$ as a function of charged-particle multiplicity although the $\rm{K}^{*0}$/K show smooth decrease as a function of charged-particle multiplicity.

\item We observe that the hadronic phase lifetime strongly depends on final state charged-particle multiplicity, system size and collision energy, while the QGP phase show very weak dependence on charged-particle multiplicity.

\item This suggests that the hadronisation from a QGP state starts at a similar temperature, which seems to be independent of charged-particle multiplicity, collision system and collision energy while the lifetime of hadronic phase is strongly dependent of final state charge-particle multiplicity. 
\item The abrupt change in behavior of kinetic freeze-out temperature and average radial flow for pp collisions after certain charged-particle multiplicity ($\simeq10-20$) suggests that the charged-particle multiplicity $\simeq10-20$ could act as minimum requirement of QGP formation.
\end{enumerate}
In view of the discovery of hints of QGP droplets in small collision systems at the LHC \cite{Sahoo:2019ifs}, this work sheds light into the role of event multiplicity on the formation of a deconfined phase and the lifetime of hadronic phase produced in various ultra-relativistic collisions. A clearer picture may emerge if the exact lifetime of QGP can be estimated for different charged-particle multiplicities, which is still an open problem. In this direction, the effective-energy that is responsible for particle production, which in principle, controls the final state multiplicity may be the responsible factor, which needs further exploration \cite{Mishra:2014dta,Sarkisyan:2015gca,Sarkisyan:2016dzo,Sarkisyan-Grinbaum:2018yld}.
\section{Acknowledgement} 
The authors acknowledge the financial supports  from  ALICE  Project  No. SR/MF/PS-01/2014-IITI(G) of Department  of  Science $\&$ Technology,  Government of India. ST and GSP acknowledge the financial support by DST-INSPIRE program of Govt. of India.

 \end{document}